\documentstyle[12pt,aaspp4]{article}

\begin{document}

\begin{flushright}
OU-TAP-49\\
\end{flushright}

\title{Critical Lines in Gravitational Lenses and the Determination 
of Cosmological Parameters}

\author{Hideki Asada}
\affil{Department of Earth and Space Science, Graduate School of Science,\\
Osaka University, Toyonaka, Osaka 560, Japan.\\}
{\it email: asada@vega.ess.sci.osaka-u.ac.jp} 

\begin{abstract}
We investigate the cosmological test recently proposed by B. Fort, 
Y. Mellier and M. Dantel-Fort (FMD), where the observed location of 
the critical line in gravitational lensing is used to 
determine the cosmological parameters, $\Omega$ and $\lambda$.  
Applying this method to the cluster of galaxies Cl0024+1654, 
FMD obtained a constraint on the cosmological constant, 
$\lambda > 0.6$, assuming the spatially flat universe. 
It plays a crucial role in this method that the angular diameter 
distance-redshift relation depends on the cosmological 
models through the cosmological parameters. 
First, using the angular diameter distance in the 
Friedmann-Lemaitre-Robertson-Walker universe, we show that one can hardly 
determine $\Omega$ by this method without the assumption of the spatially 
flat universe. 
We also investigate the effect of inhomogeneities of the universe 
by using the Dyer-Roeder angular diameter distance. 
It is shown that the effect of inhomogeneities can become 
too large to be ignored, particularly for a high density universe. 
As a result, this method cannot be taken as a clear cosmological test 
to determine $\Omega$ and $\lambda$, though it may provide a bound on 
$\Omega$ and $\lambda$. 
Moreover, we mention the uncertainty of the determination of 
the velocity dispersion, which is regarded as one of the most serious 
problems in this test. 
\end{abstract}


\newpage
\baselineskip 8mm

\section{Introduction}

In understanding the Universe, it has been one of central issues to determine 
the cosmological parameters; the density parameter $\Omega$ and 
the cosmological constant $\lambda$ (Weinberg 1972; Peebles 1993). 
Although some cosmological tests have been proposed and applied, 
the conclusive result has not been obtained yet. 
The difficulties in determining the cosmological parameters consist in 
the following fact: All cosmological tests are based on the fact that 
relations among observables depend on the cosmological parameters. 
However, the relations also rely on astrophysical models. 
For instance, the galaxy-number count is significantly affected by the 
formation and evolution of galaxies (Yoshii and Takahara 1988; 
Fukugita, Takahara, Yamashita and Yoshii 1990), and the statistics 
of multiple images of quasars due to the gravitational lensing 
depends greatly on the lens model of galaxies and cluster 
of galaxies (Flores and Primack 1996; Kochanek 1996). 

The method to determine the deceleration parameter $q_0$, taking type Ia 
supernovae (SNIa) as the standard candle, has been proposed. 
Recently, it became possible to apply the method to observational data, 
because of a progress of the search for high redshift SNIa 
(Perlmutter et.al. 1995, 1996; references therein). 
This method makes use of the dependence of the luminosity distance-redshift 
relation on cosmological parameters. 
It is assumed that the mean absolute magnitude of SNIa 
is constant over various redshifts. 
The intrinsic dispersion of magnitude of SNIa can be taken into account 
in the statistical sense. 
Nevertheless, it is likely that the gravitational lensing effect due to 
inhomogeneities of the universe disturbs this test, which has recently been 
investigated by several authors using various approaches 
(Kantowski, Voughan and Branch 1995; Wambsganss et.al. 1996; Frieman 1996). 

On the other hand, the angular diameter distance-redshift relation 
also depends on cosmological parameters. 
Thus, it is possible to determine cosmological parameters through the 
dependence, if one can measure the redshift and the apparent size of 
an object whose physical size is known by another independent method. 
B. Fort, Y. Mellier and M. Dantel-Fort have taken notice of the relation 
between the mass of the gravitational lens (cluster of galaxies) 
and the location of the critical line (Fort, Mellier and Dantel-Fort 1996; 
hereafter FMD).  
Using the data of cluster of galaxies Cl0024+1654, they obtained 
a constraint $\lambda > 0.6$, as a preliminary result. 
Their procedure is as follows (See section 2 for more details): 
{}From the observation of a lens object (which is nearly axisymmetric 
along a line of sight) and background galaxies, one can determine 
the velocity dispersion, the angular core radius and 
the maximum angular radius of critical lines. 
In particular, the critical line, defined as the location where 
large amplification of images occurs (Schneider, Ehlers and Falco 1992, 
hereafter SEF), can be determined observationally as the location 
where the number density of background galaxies is much larger than 
that in the field (FMD). 
The location of the critical line depends not only on the mass distribution 
of the lens but also on the ratio between two distances; 
the angular diameter distance from the lens to the source and 
that from the observer to the source. 
The mass distribution of the lens is usually estimated from the velocity 
dispersion and the core radius. 
When the redshift of the source is between 3 and 5, the distance ratio 
depends strongly on the cosmological parameters but is insensitive to 
the redshift of the source. 
Thus the cosmological parameters can be determined by this dependence. 
FMD have studied the case of the spatially flat universe. 
However, it is still important to investigate, without assuming the 
spatially flat universe, how $\Omega$ and $\lambda$ can be really 
constrained by this method. 
Moreover, the inhomogeneities of the universe may disturb this test by 
bringing uncertainties to the angular diameter distance-redshift relation. 
We also investigate this effect by using the Dyer-Roeder (DR) angular
diameter distance (Dyer and Roeder 1972, 1974;hereafter DR). 

This paper is organized as follows. 
In section 2, the method by FMD is explained. 
In section 3, we consider cases of general $\Omega$ and $\lambda$ 
in order to investigate how accurately $\Omega$ and $\lambda$ can be 
determined by this method.
In section 4, the DR angular diameter distance is used in order to 
take into consideration the effect of clumpiness of the universe. 
In section 5, we discuss how $\Omega$ and $\lambda$ can be 
constrained in the clumpy universe. 
We also mention the uncertainty of the determination of 
the velocity dispersion. 
Section 6 is devoted to summary.

\section{The location of the critical lines and cosmological parameters}

In this section, we would like to summarize the method to determine 
the cosmological parameters by the use of the maximum radius of 
the critical lines in the gravitational lens. 
It is assumed that the mass profile of the cluster of galaxies is that of 
the isothermal sphere with core (FMD). 
Then the mass profile of the cluster is written as 
\begin{equation}
\rho(r)=\rho_0 \Bigl(1+\frac{r^2}{r_c^2}\Bigr)^{-1} , \label{densityiso}
\end{equation}
where $r_c$ is a core radius, $\rho_0=\sigma^2/2\pi G r_c^2$ and 
$\sigma$ is a line-of-sight velocity dispersion. 
The surface mass density projected to the lens plane is useful in 
investigating gravitational lensing and obtained from 
Eq.(\ref{densityiso}) as 
\begin{equation}
\Sigma(R)=\frac{\sigma^2}{2G}\frac1{\sqrt{R^2+r_c^2}} , 
\end{equation}
where $R$ is a radius from the center of the cluster on the lens plane. 
The mass within the radius $R$ on the lens plane is expressed as 
\begin{equation}
M(R)={\pi \sigma^2 \over G} \Bigl( \sqrt{R^2+r_c^2}-r_c \Bigr). \label{massR}
\end{equation}
In the spherically symmetric case such as the isothermal sphere, 
the critical line appears on the Einstein ring (SEF); 
\begin{equation}
\theta_{cr}=\theta_E=\sqrt{\frac{4GM}{c^2}\frac{D_{LS}}{D_{OL}D_{OS}}} , 
\label{thetacr}
\end{equation}
where 
$\theta_{cr}$ and $\theta_E$ are the angular radius of the critical line 
and that of the Einstein ring, and $D_{OL}$, $D_{OS}$ and $D_{LS}$ 
are the angular diameter distances between the observer and the lens, 
between the observer and the source, and between the lens and 
the source, respectively. 
In the case of isothermal sphere, we may use Eq.$(\ref{massR})$ to obtain 
\begin{equation}
\theta_{cr}=\sqrt{ \frac{4\pi \sigma^2}{c^2}\Bigl((R_{cr}^2+r_c^2)^{1/2}
-r_c\Bigr) \frac{D_{LS}}{D_{OL}D_{OS}} } , \label{isocr}
\end{equation}
where $R_{cr}$ is the radius of the critical line. 
It should be noted that the observable is not the radius of the 
critical line but its angular radius, $\theta_{cr}=R_{cr}/D_{OL}$. 
{}From Eq.($\ref{isocr}$), we obtain the important ratio (FMD), 
which depends only on the observables, 
\begin{equation}
\frac{D_{LS}}{D_{OS}}=\frac{c^2}{4\pi \sigma^2}\Bigl((\theta_{cr}^2
+\theta_c^2)^{1/2}+\theta_c \Bigr) , \label{isocritical}
\end{equation}
where we defined $\theta_c=r_c/D_{OL}$. 
Usually, the radius of the critical line (several tens arcsec, several Mpc) 
is much larger than the core radius (less than Mpc). 
Then we can use the approximate relation 
\begin{equation}
\frac{D_{LS}}{D_{OS}}\approx\frac{c^2}{4\pi \sigma^2} \theta_{cr} . 
\label{isocritical2}
\end{equation}

It is noteworthy that the angular diameter distance has not been specified. 
Since the angular radius of the critical line $\theta_{cr}$, 
the line-of-sight velocity dispersion in the cluster $\sigma$ and 
the angular core radius $\theta_{c}$ can be obtained observationally, 
the ratio $D_{LS}/D_{OS}$ can be determined from these data. 
On the other hand, the distance ratio $D_{LS}/D_{OS}$ as well as 
the angular diameter distances, $D_{LS}$ and $D_{OS}$, depends on 
the cosmological models. 
For instance, in the FLRW universe, the relation between the redshift and 
the distance ratio depends only on the cosmological parameters, 
$\Omega$ and $\lambda$ (Weinberg 1972; Peebles 1993). 
Figure 1 shows how the relation between the distance ratio $D_{LS}/D_{OS}$ 
and the redshift of the source depends on the cosmological parameters, 
where we assume the angular diameter distance in the FLRW universe. 
It also shows that $D_{LS} / D_{OS}$ increases with the redshift of 
the source. 
It should be noted in Fig.1 that $D_{LS}/D_{OS}$ is sensitive to $\Omega$ 
and $\lambda$, but insensitive to the redshift of the source from 3 to 5. 
Therefore, if one obtains observationally the maximum of $D_{LS} / D_{OS}$, 
the cosmological parameters can be constrained without the precise 
measurement of redshift of background faint galaxies (FMD). 
In order to constrain the cosmological parameters, it is necessary 
to obtain $D_{LS}/D_{OS}$ within about ten percent accuracy, 
for Cl0024+1654 at $z_L=0.39$, as shown by Fig.1. 

{}From Eq.($\ref{thetacr}$), it is clear that the cosmological test 
discussed here relies on (1) which angular diameter distance to be used 
and (2) the mass of the cluster as the gravitational lens. 
We mainly investigate the effect of the first aspect 
on the cosmological test. In particular, we pay much attention to 
the distance ratio $D_{LS}/D_{OS}$. 
It is worthwhile to mention that the cosmological test discussed here 
does not depend on the Hubble constant because of our using only 
the distance ratio $D_{LS}/D_{OS}$.

\section{Cosmological test in the FLRW universe}

In the previous section, we have given Fig.1, which shows a relation 
among $z_S$, $D_{LS}/D_{OS}$ and cosmological parameters, by assuming 
$\Omega+\lambda=1$. 
In the remaining of this paper, we fix the redshift of the source 
at $z_S=4$ in order to make our discussion clear. 
This treatment is plausible in practice, since $D_{LS}/D_{OS}$ is 
insensitive to the redshift of the source for $z_S >3$, as shown by Fig.1. 
Figure 2 shows relations between $\Omega$ and $D_{LS}/D_{OS}$ 
in two cases of $\Omega+\lambda=1$ and $\lambda=0$.  
It is found that a bound on $\lambda$ can be obtained in the case of 
$\Omega+\lambda=1$, particularly for large $D_{LS}/D_{OS}$. 
Remarkably, $D_{LS}/D_{OS}$ depends very weakly on $\Omega$ 
when $\lambda=0$. 
However, it is not clear whether $D_{LS}/D_{OS}$ is generally 
insensitive to $\Omega$. 
Thus we investigate the dependence of $D_{LS}/D_{OS}$ on 
$\Omega$ and $\lambda$. 
In order to investigate a relation among $\Omega$, $\lambda$ and 
$D_{LS}/D_{OS}$, we plot contours of $D_{LS}/D_{OS}$ on 
$\Omega$-$\lambda$ plane assuming $z_L=0.5$, as shown by Fig.3. 
Figure 3 shows clearly that $D_{LS}/D_{OS}$ is very insensitive to 
$\Omega$ regardless of $\lambda$. 
Therefore, it is difficult to determine $\Omega$ by this method. 
However, when we assume the spatially flat universe, one can put a 
bound on the cosmological constant, as FMD discussed. 
We use Fig.3, in order to illustrate how we can constrain the cosmological 
constant for a given $D_{LS}/D_{OS}$ assuming $\Omega+\lambda=1$. 
For instance, $D_{LS}/D_{OS} \sim 0.80$ would suggest a universe dominated 
by the cosmological constant ($\lambda \sim 0.9$). 
On the other hand, if $D_{LS}/D_{OS}$ is less than $0.67$, 
the vanishing cosmological constant is preferred. 

Figs.4(a)-(c) show contours of $D_{LS}/D_{OS}$ on the 
$\Omega$-$\lambda$ plane in the cases of $z_L=0.1, 0.3$ and $1.0$ 
respectively . 
These suggest that, regardless of the redshift of the lens, $D_{LS}/D_{OS}$ 
depends very weakly on $\Omega$, but is more sensitive to $\lambda$ 
in a lower density universe. 
Thus, one may constrain the cosmological constant by using the lens 
at various redshifts as well. 
However, as shown by Fig.4(a), the cosmological constant does not change 
$D_{LS}/D_{OS}$ so much for a low redshift lens, that one must use a lens 
at a moderately high redshift ($z_L > 0.3$) for this cosmological test. 
It is important to note the difference between the new method and 
the so-called $q_0$ test; the new method is more sensitive to $\lambda$ 
than to $\Omega$, while the $q_0$ test aims at the determination of 
the combination $\Omega/2-\lambda$ in the matter-dominated 
universe (Weinberg 1972; Peebles 1993).

\section{Cosmological test in the clumpy universe} 

On a sufficiently large scale, our universe can be described by 
the FLRW universe. 
Nevertheless, the light ray propagates from the source to the observer 
according to the geodesic equation which is determined by the local metric 
along the light ray (Weinberg 1972; Peebles 1993). 
Therefore, it is likely that the light propagation is affected by the 
inhomogeneities of the universe. 
Much attention has focused on the effect of the inhomogeneities 
on the luminosity or angular diameter distance between the observer 
and the source, $D_{OS}$ (DR; Futamase and Sasaki 1989; 
Kasai, Futamase and Takahara 1990; SEF; Sasaki 1993). 
Thus, here we must consider how much the inhomogeneities affect 
the ratio of angular diameter distances $D_{LS}/D_{OS}$. 

We consider the DR angular diameter distance, 
for its simplicity, in order to take into account 
the inhomogeneities (DR; SEF). 
The DR angular diameter distance is determined by 
\begin{equation}
{d^2 \over dw^2}D+{3 \over 2}(1+z)^5 \alpha \Omega D=0 , 
\label{raychaudhuri}
\end{equation}
where the parameter $\alpha$ represents the clumpiness along the light ray. 
In the FLRW universe, $\alpha$ is unity, while $\alpha$ vanishes 
when the light ray propagates through the empty space. 
In this paper, we consider $0\leq\alpha\leq 1$, since we have little 
knowledge of $\alpha$. 
Here $w$ is an affine parameter, which is assumed (SEF) to be given by  
that in the FLRW universe 
\begin{equation}
{dz \over dw}=(1+z)^2 \sqrt{ \Omega z (1+z)^2 -\lambda z (2+z)+(1+z)^2 } . 
\label{affine}
\end{equation}
It is noteworthy that the coefficient of the last term of 
Eq.($\ref{raychaudhuri}$), $3\alpha\Omega/2$, comes from the Ricci 
focusing by the clumpy matter and, therefore, the DR angular diameter 
distance is a decreasing function of $\alpha$ for a fixed 
redshift (DR; SEF). 

Since we have considered the FLRW universe ($\alpha=1$) in the previous 
section, here we consider the other limiting case $\alpha=0$ in order to 
investigate the effect of clumpiness of the universe. 

\subsection{The dependence on $\Omega$ and $\lambda$}

By solving Eq.($\ref{raychaudhuri}$) numerically, we can obtain 
the angular diameter distances $D_{LS}$ and $D_{OS}$. 
As a result, we draw Fig.5 which shows relations between $\Omega$ and 
$D_{LS}/D_{OS}$ in two cases of $\Omega+\lambda=1$ and $\lambda=0$. 
It is found that, in the clumpy universe,  the distance ratio 
$D_{LS}/D_{OS}$ is smaller than that in the case of $\alpha=1$. 
In particular for a higher density universe, the dependence of 
$D_{LS}/D_{OS}$ on $\alpha$ is larger in both cases; 
$\Omega+\lambda=1$ and $\lambda=0$. 
Next, we draw contours of $D_{LS}/D_{OS}$ on $\Omega$-$\lambda$ plane 
in Fig.6, in order to clarify the dependence of $D_{LS}/D_{OS}$ 
on $\Omega$ and $\lambda$ in the clumpy universe. 
By comparing Fig.6 with Figs.3 and 4, we find that, in the universe 
with any $\Omega$ and $\lambda$, $D_{LS}/D_{OS}$ for $\alpha=0$ is 
smaller than that for $\alpha=1$. 
For any $\lambda$, we also find that the discrepancy of $D_{LS}/D_{OS}$ 
between $\alpha=0$ and $1$ increases with $\Omega$. 
This may be attributed to the fact that the focusing effect due to the 
clumpiness in Eq.($\ref{raychaudhuri}$) increases with $\Omega$. 
Furthermore, Fig.6 implies that, for $\alpha=0$,  $D_{LS}/D_{OS}$ is 
very sensitive to $\Omega$ rather than to $\lambda$. 
This is in contrast to the case of $\alpha=1$. 
We will show an important property of $D_{LS}/D_{OS}$ in the next 
subsection, before we will discuss in section 5 how one can constrain 
the cosmological parameters by taking into account the effect due to 
the clumpy universe.

\subsection{The general case} 

In the above, we have only considered two limiting cases of 
$\alpha=0$ and $1$. 
In Figs.2 and 5, the following interesting property of 
the distance ratio appears 
\begin{equation}
{ D_{LS} \over D_{OS} }(\alpha=0;\mbox{Empty}) < 
{ D_{LS} \over D_{OS} }(\alpha=1; \mbox{FLRW}) . \label{inequality}
\end{equation}

By noticing that $\alpha$ enhances monotonically the Ricci focusing in 
Eq.($\ref{raychaudhuri}$), one expects that Eq.$(\ref{inequality})$ is 
generalized to a relation for $\alpha_1 < \alpha_2$ as 
\begin{equation}
{ D_{LS} \over D_{OS} }(\alpha_1) < 
{ D_{LS} \over D_{OS} }(\alpha_2) . \label{inequality2}
\end{equation}
This is proved as follows: 
Let us fix the redshift of the source and the cosmological parameters, 
$\Omega$ and $\lambda$. 
Then we define a function of $z_L$ parameterized by $\alpha$ as 
\begin{equation}
X_{\alpha}(z_L)={D_{LS} \over D_{OS}}(\alpha) . 
\end{equation}
From Eq.$(\ref{raychaudhuri})$, we obtain the equation 
for $X_{\alpha}(z_L)$ 
\begin{equation}
{d^2 \over dw_L^2}X_{\alpha}(z_L)+{3 \over 2}(1+z_L)^5 \alpha \Omega 
X_{\alpha}(z_L)=0 , \label{raychaudhuri2}
\end{equation}
where $w_L$ is an affine parameter at the lens. 
We define the Wronskian as 
\begin{equation}
W(X_{\alpha_1},X_{\alpha_2})=\Bigl( X_{\alpha_1}{d X_{\alpha_2} \over 
d w_L}-X_{\alpha_2}{d X_{\alpha_1} \over d w_L} \Bigr) . 
\end{equation}
Using Eq.$(\ref{raychaudhuri2})$, we obtain 
\begin{equation}
{d \over d w_L}W(X_{\alpha_1},X_{\alpha_2}) < 0 , \label{wronskian2}
\end{equation}
where we used $\alpha_1 < \alpha_2$. 
Since both $X_{\alpha_1}$ and $X_{\alpha_2}$ vanish at $z_L=z_S$, 
we obtain 
\begin{equation}
W(X_{\alpha_1}(z_S),X_{\alpha_2}(z_S))=0 . \label{wronskian3} 
\end{equation}
From Eqs.$(\ref{wronskian2})$ and $(\ref{wronskian3})$, we find 
\begin{equation}
W(X_{\alpha_1},X_{\alpha_2}) >0 , \label{wronskian4}
\end{equation}
where we used the fact that the affine parameter $w$ defined 
by Eq.$(\ref{affine})$ is an increasing function of $z$. 
Eq.($\ref{wronskian4}$) is rewritten as 
\begin{equation}
{d \over d w_L} \ln{X_{\alpha_2} \over X_{\alpha_1}} > 0 . 
\label{wronskian5}
\end{equation}
Since $X_{\alpha}$ always becomes unity at the observer, we find 
\begin{equation}
\ln{X_{\alpha_1}(z_L=0) \over X_{\alpha_1}(z_L=0)}=0 . \label{wronskian6}
\end{equation}
From Eqs.$(\ref{wronskian5})$ and $(\ref{wronskian6})$, we find 
\begin{equation}
\ln{X_{\alpha_2} \over X_{\alpha_1}} >0 . \label{proof}
\end{equation}
This gives us what we wish to prove. 
It should be noted that Eq.($\ref{proof}$) holds even if one uses 
the opposite sign in the definition of the affine parameter 
in Eq.$(\ref{affine})$. 
Eq.($\ref{inequality2}$) means that the distance ratio $D_{LS} / D_{OS}$ 
for $0<\alpha<1$ takes a value between those for $\alpha=0$ and $1$, 
namely for $0<\alpha<1$, 
\begin{equation}
{ D_{LS} \over D_{OS} }(\alpha=0;\mbox{Empty}) < 
{ D_{LS} \over D_{OS} }(\alpha) <
{ D_{LS} \over D_{OS} }(\alpha=1; \mbox{FLRW}) . \label{inequality3}
\end{equation}
Therefore, all we have to do is to investigate two limiting cases 
$\alpha=0$ and $1$, in order to know a region where $D_{LS} / D_{OS}$ 
for $0 \leq\alpha\leq 1$ takes its value for given $z_L$, $\Omega$ 
and $\lambda$. 
Eq.($\ref{inequality3}$) shows that, for the light propagation through 
a lower density region such as a void, the distance ratio $D_{LS} / D_{OS}$ 
is always less than that in the FLRW universe.

\section{Discussion}
We have used the DR angular diameter distance, in order to 
investigate how much the clumpy universe affects the determination of the 
cosmological parameters by the method using the location of the critical 
lines in the gravitational lens. 
It is shown in section 4 that the distance ratio $D_{LS}/D_{OS}$ 
in a clumpy universe is necessarily less than that in the FLRW universe. 
It is also shown that the distance ratio $D_{LS} / D_{OS}$ in a higher 
density universe is more affected by the clumpiness. 
As a result, it seems impossible that the cosmological parameters 
can be clearly determined by this method. 
This situation may remind us of the recent discussion about the $q_0$ test 
using the luminosity distance-redshift relation, where 
about fifty-percent deviation can happen in the Swiss cheese model 
of the universe (Kantowski, Vaughan and Branch 1995). 

Nevertheless, there are some cases where the cosmological parameters 
can be constrained by noticing that the clumpiness always decreases 
$D_{LS}/D_{OS}$. 
In order to illustrate these cases, we use Figs.3 and 6(c) ($z_L=0.5$), 
assuming two cosmological models with (1) $\Omega+\lambda=1$ and 
(2) $\lambda=0$.  
First, we consider the case of (1) $\Omega+\lambda=1$: 
In the case of $D_{LS}/D_{OS}$ between 0.67 and 0.85, one can put 
a bound on $\lambda$, though the bound is sensitive to $\alpha$. 
The allowed region of $\lambda$ becomes wide as $D_{LS}/D_{OS}$ decreases. 
If $D_{LS}/D_{OS}$ is between 0.53 and 0.67, an upper bound on $\lambda$ 
can be obtained, which is given by $\alpha=0$. 
Next, we use the case of (2) $\lambda=0$: 
In the case of $D_{LS}/D_{OS}>0.63$, one can put a lower bound on $\Omega$, 
which is given by $\alpha=1$. 
It is noteworthy that $\Omega$ must be greater than unity 
if $D_{LS}/D_{OS}$ exceeds $0.67$. 
If $D_{LS}/D_{OS}$ is less than 0.63, one can also put a lower bound 
on $\Omega$, which is given by $\alpha=0$. 
In the above, we assumed either $\Omega+\lambda=1$ or $\lambda=0$ a priori. 
However, even without such a constraint, the allowed region of the 
parameters $\Omega$ and $\lambda$ is still bounded by two limiting cases; 
$\alpha=0$ and $1$ (For instance, see two curves for $D_{LS}/D_{OS}=0.69$ 
in Fig.6(c)). 
It is important to mention that a lower bound on $\lambda$, given by 
using the angular diameter distance in the FLRW universe ($\alpha=1$), 
is not affected by the clumpiness. 
Thus, $\lambda >0.6$ given by FMD cannot be relaxed by taking into account 
the clumpiness of the universe. 
Here, it should be noted that the cosmological constant may be well 
constrained by this method, if the density parameter can be determined 
by another method. 

In the above, we have considered only the dependence of $D_{LS}/D_{OS}$ 
on the cosmological parameters. 
Here, we briefly mention the dependence of this test on astrophysical models 
of the cluster of galaxies. 
As shown in Eq.$(\ref{isocritical})$, the precise determination of 
the velocity dispersion is necessary even if one assumes the isothermal 
sphere as the lens. 
If one specifies the redshifts of the lens and the source, the accuracy 
required for the determination of the velocity dispersion depends 
on which cosmological models we wish to compare with. 
If one wishes to distinguish, for instance, the cosmological model 
with $(\Omega, \lambda)=(0.2, 0)$ from that with $(\Omega, \lambda)
=(0.2, 0.8)$, one must distinguish $D_{LS}/D_{OS}=0.64$ from $0.76$ 
for $z_L=0.5$ and $z_S=4$. 
This means that one must discriminate the difference of 
$D_{LS}/D_{OS}$ at worst by 16 percent. 
However, the statistical error of the velocity dispersion is at least 
about ten percent because of the limited number(about a hundred) of 
galaxies inside the radius of the critical line, 
even though the velocity of individual galaxies can be measured 
precisely (Dressler, Gunn and Schneider 1985). 
This required accuracy is almost unchanged around $z_L=0.5$. 
Thus, because of the statistical error of the determination of
the velocity dispersion, it is concluded that the current data of Cl0024+1654
does not rule out the cosmological model without $\lambda$.
However, in the statistical sense, it is possible to put a bound on 
$\Omega$ and $\lambda$, if one can apply this method to 
a lot of clusters of galaxies. 
For instance, if one uses ten clusters of galaxies, the statistical error 
of the velocity dispersion becomes less than four percent, so that one may 
distinguish some cosmological models, say with $(\Omega, \lambda)=(0.2, 0)$ 
and $(0.2, 0.8)$. 

Furthermore, the mass profile used as the lens model must be confirmed 
in order to make this cosmological test reliable. 
Thus it is of great importance that the mass or the mass profile of the lens 
is determined precisely.  
It is hoped that the mass reconstruction by the weak gravitational lensing 
(Kaiser and Squires 1993; Kaiser 1995; Seitz and Schneider 1995a, 1995b) 
as well as the analysis of the velocity field in the cluster of galaxies 
(Yee, Ellingson and Carlberg 1996) clarifies the precise mass profile of 
the cluster as the lens. 
Then the cosmological test discussed here will work well.

\section{Summary}

Without assuming the spatially flat universe, we have reexamined 
the cosmological test done by FMD, by using the DR angular diameter distance 
in order to take into account the clumpiness of the universe. 
If the angular diameter distance is that in the FLRW universe, namely 
ignoring the effect of the clumpiness, it is found that $\Omega$ cannot 
be determined directly by this method. 
It was also shown that the effect of inhomogeneities of the universe 
can become too large to be ignored, particularly for a high density universe. 
Therefore, this method cannot be taken as a clear cosmological test 
to determine $\Omega$ and $\lambda$. 
However, it is possible to provide a bound on $\Omega$ and $\lambda$ 
in the clumpy universe. 
For more quantitative discussion, it is necessary to consider 
the gravitational lensing in a more realistic inhomogeneous universe. 

In section 5, we mentioned that the statistical error of the velocity 
dispersion can be the most serious problem in the cosmological test 
discussed here. 
Therefore, in order to constrain $\Omega$ and $\lambda$ by this method, 
it is necessary to observe many clusters of galaxies. 
For instance, Sloan Digital Sky Survey (SDSS) will provide us 
with a lot of clusters of galaxies which are useful 
for this cosmological test.

\bigskip

\acknowledgements

The author would like to thank M. Sasaki and T. Tanaka for careful 
reading of the manuscript and invaluable comments on it. 
He is also very grateful to B. F. Roukema for pointing out the recent 
work by B. Fort, Y. Mellier and M. Dantel-Fort. 
He would like to thank S. Ikeuchi and M. Sasaki 
for useful advice and continuous encouragement.




\newpage

\figcaption[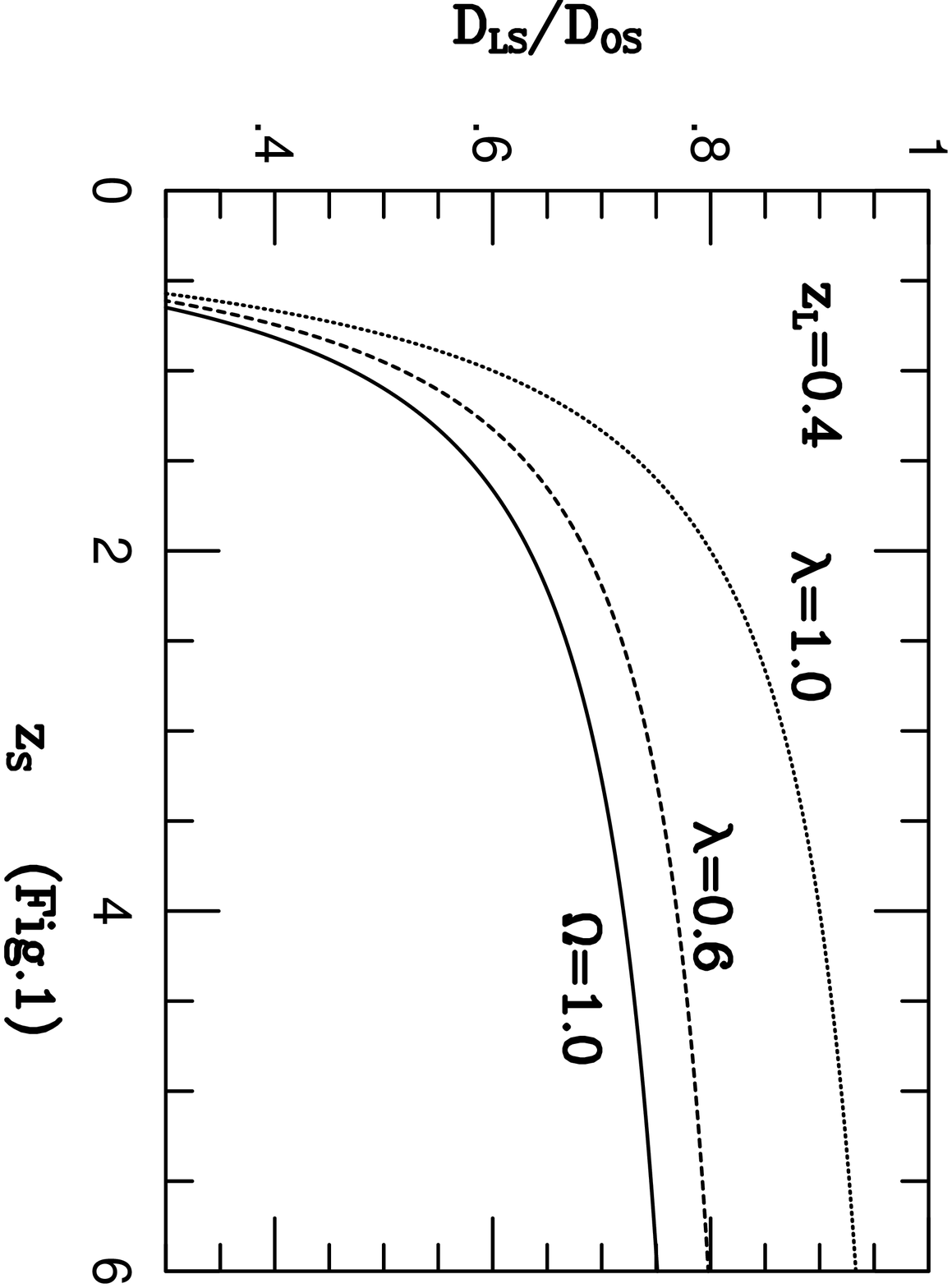]{
The distance ratio $D_{LS}/D_{OS}$ for the lens at $z_L=0.4$. 
$(\Omega,\lambda)=(1,0)$, $(\Omega,\lambda)=(0.4,0.6)$ and 
$(\Omega,\lambda)=(0,1)$ are shown by solid, dashed and dotted lines 
respectively. 
}

\figcaption[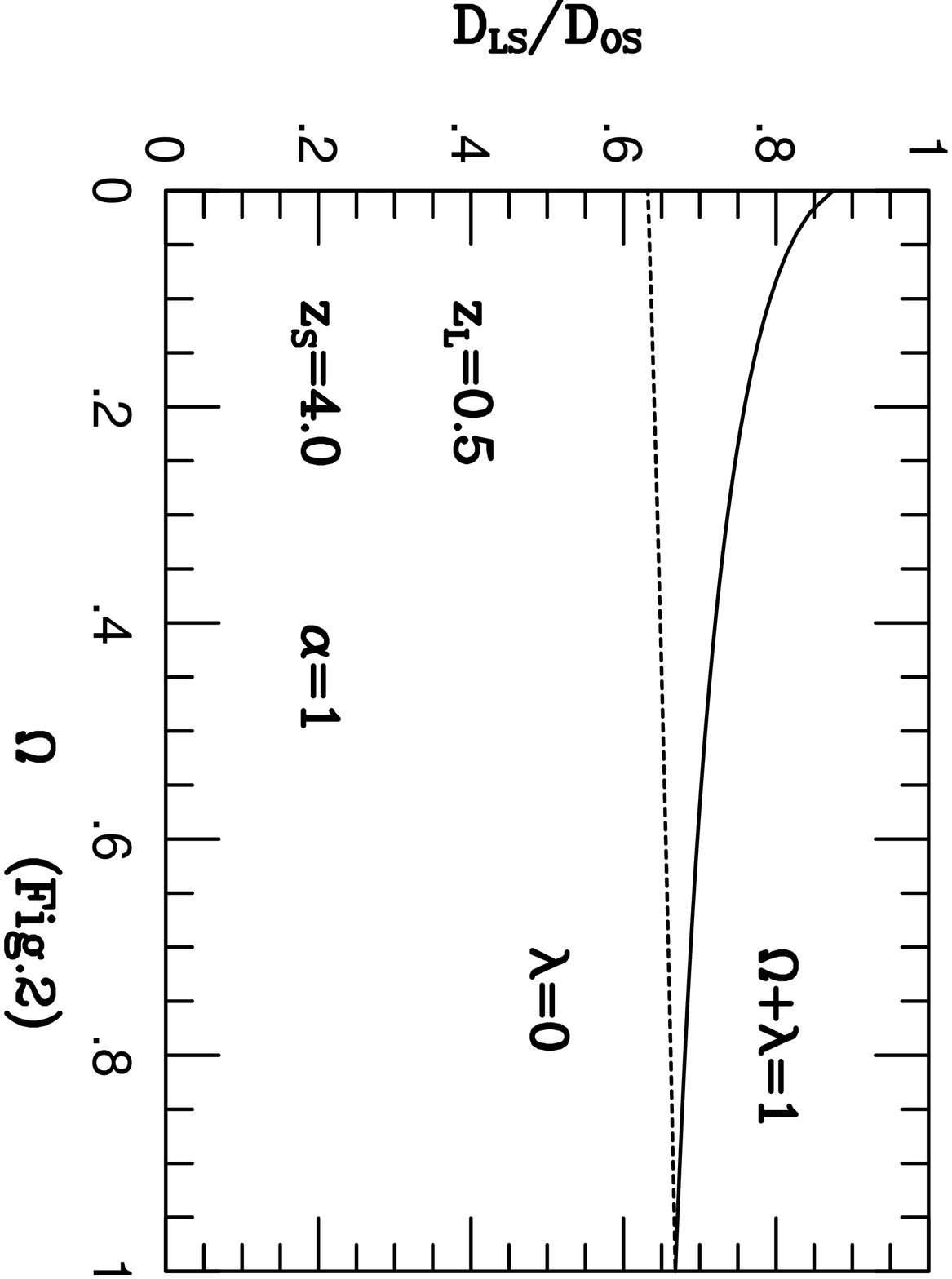]{
The distance ratio $D_{LS}/D_{OS}$ for the lens at $z_L=0.5$ 
and the source at $z_S=4$. 
We use the angular diameter distance in the FLRW universe. 
We assume $\Omega+\lambda=1$ (solid line) and $\lambda=0$ (dashed line). 
}

\figcaption[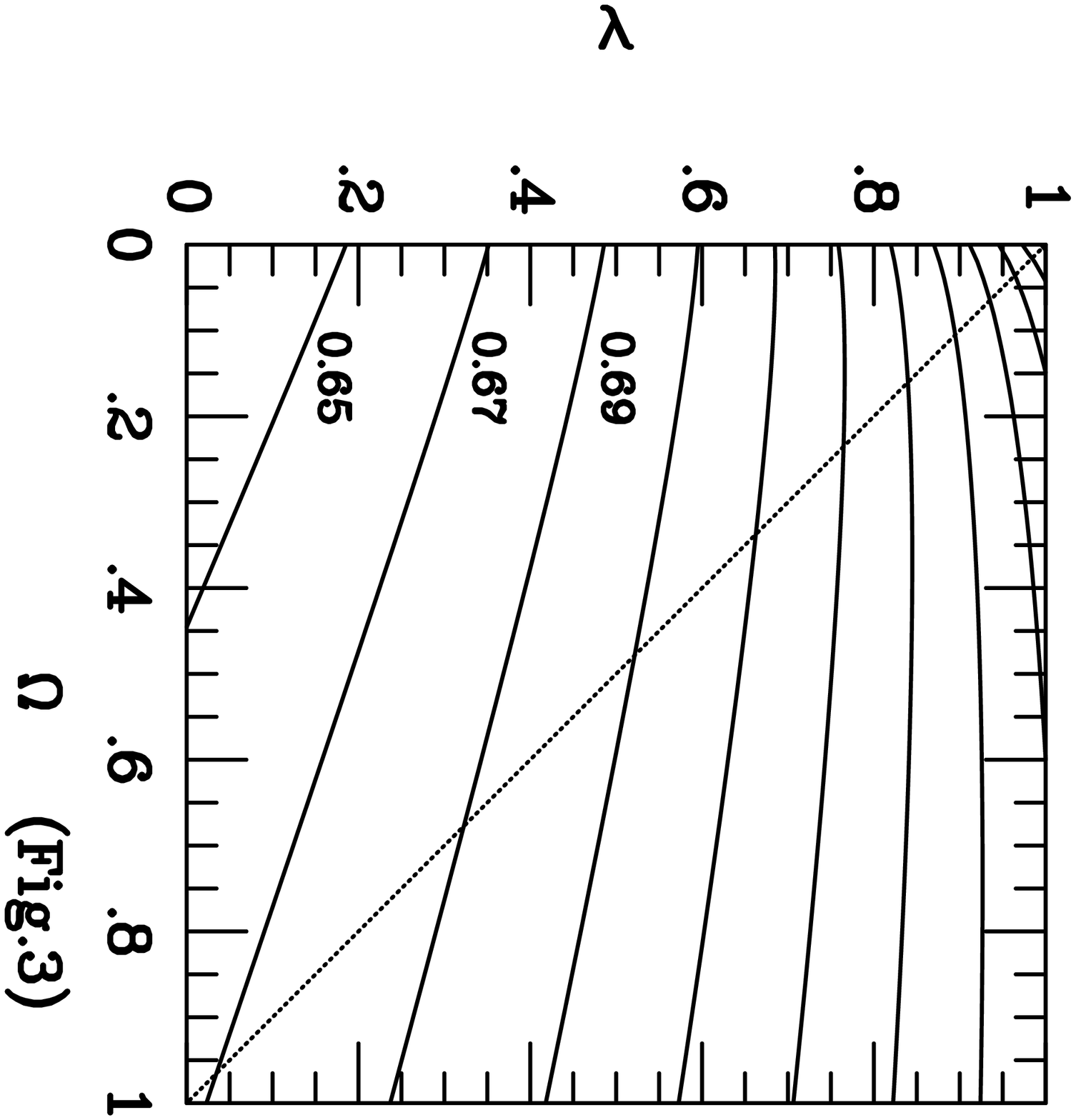]{
Contours of $D_{LS}/D_{OS}$ on $\Omega$-$\lambda$ plane 
Each contour from top to bottom is drawn where $D_{LS}/D_{OS}$ runs 
from $0.85$ to $0.65$ at the interval of $0.02$. 
The dotted line connecting $(\Omega, \lambda)=(1,0)$ and $(0,1)$ 
indicates the spatially flat universe. 
}

\figcaption[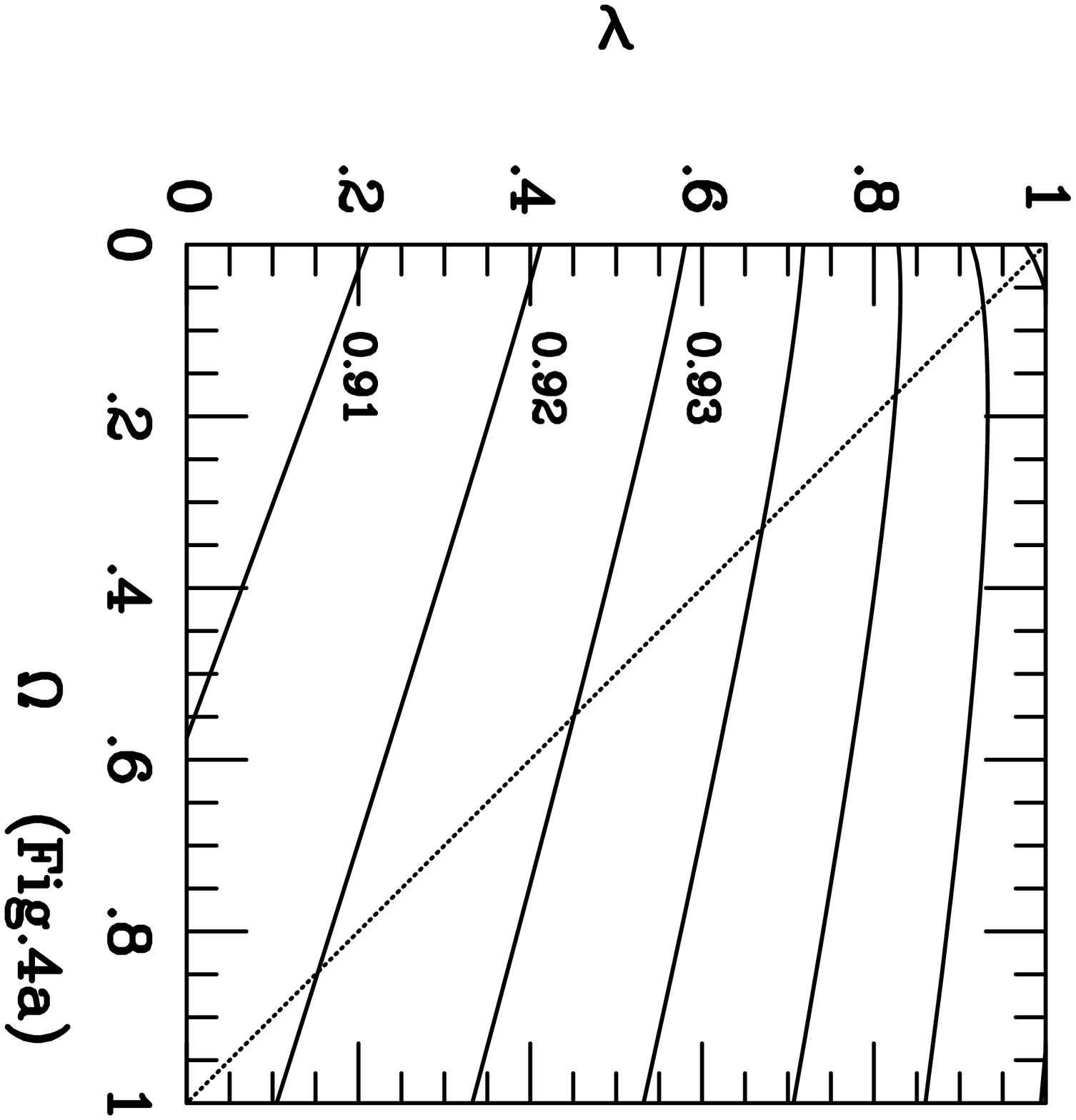,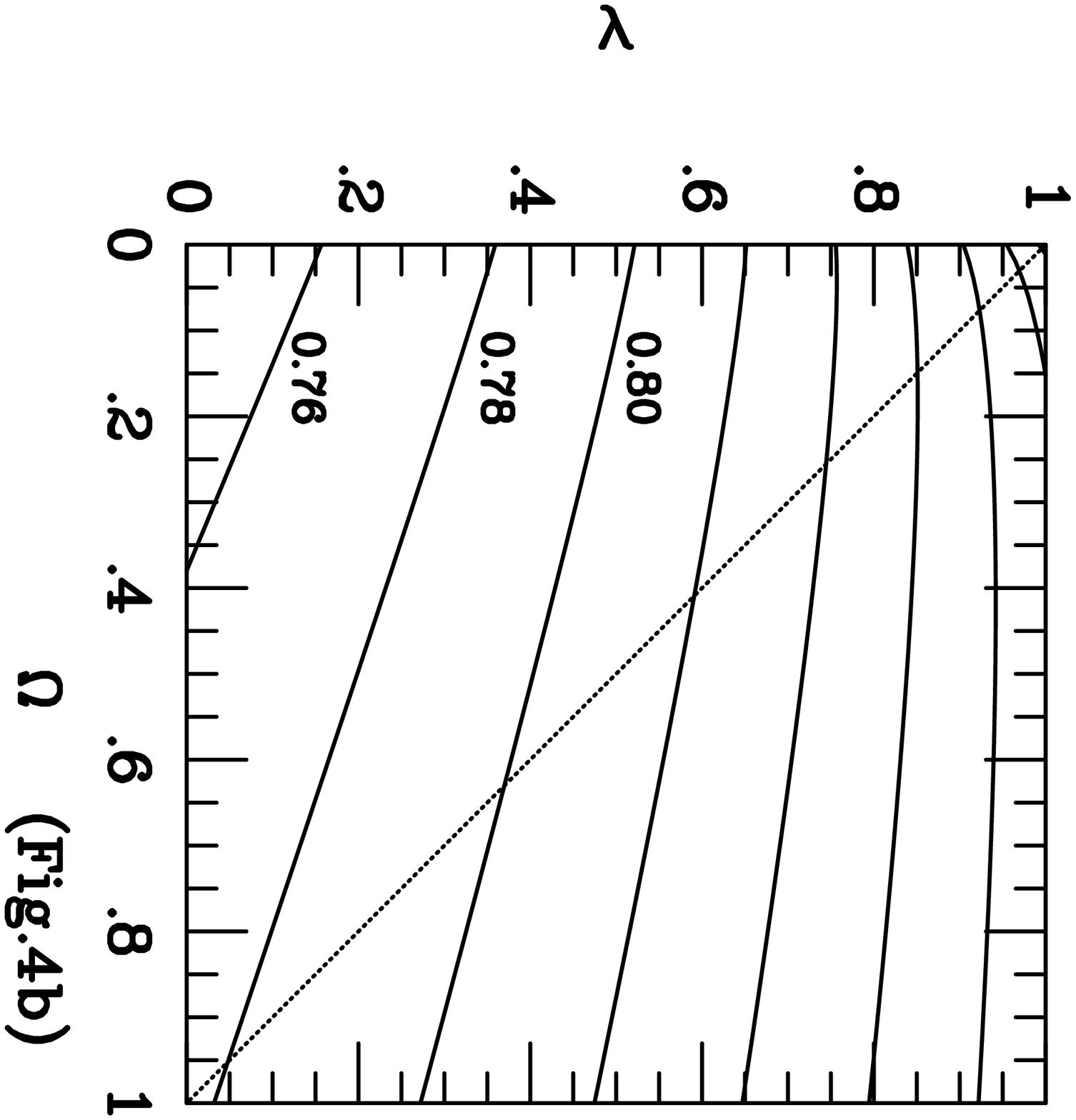,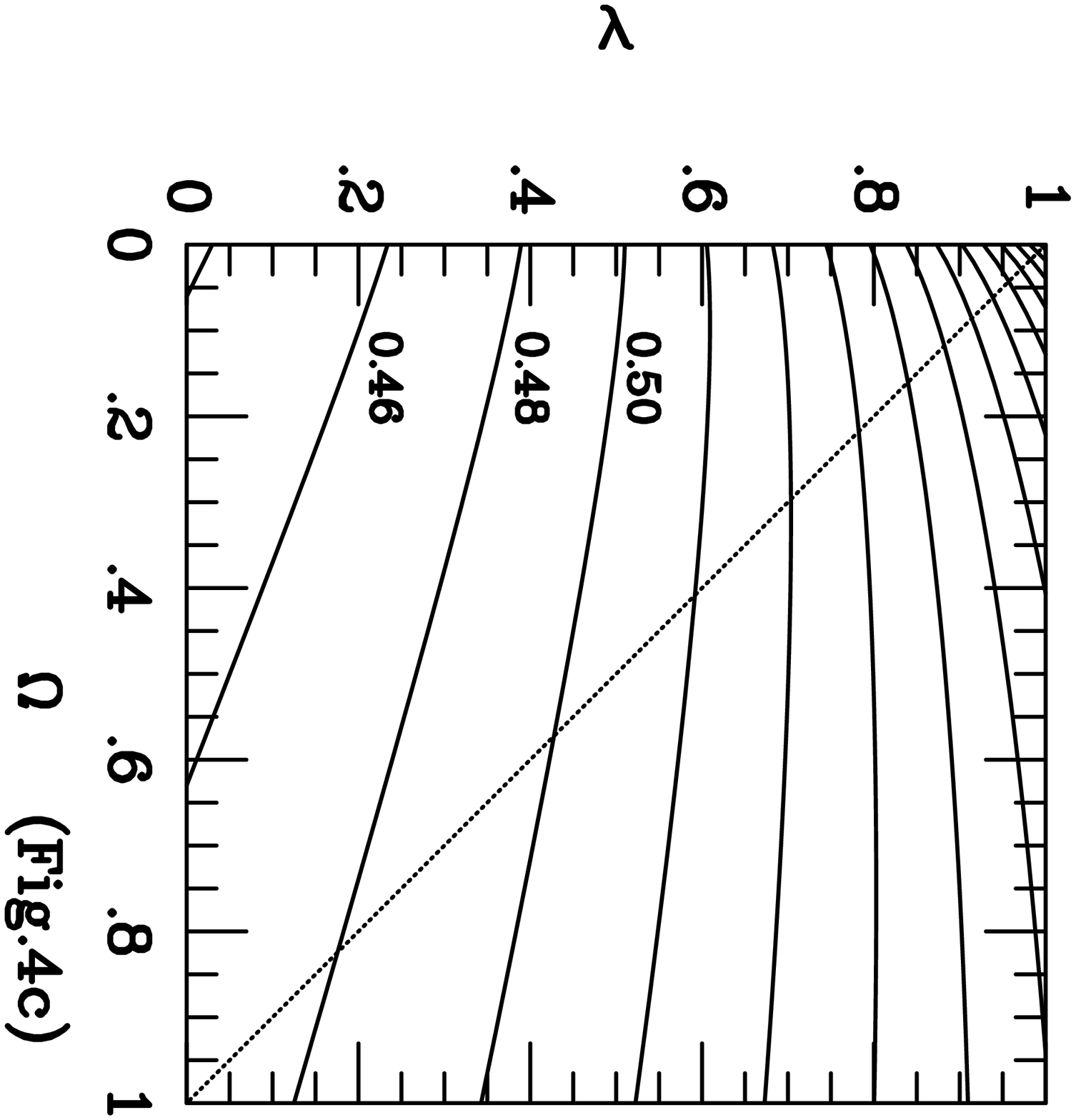]{
Contours of $D_{LS}/D_{OS}$ on $\Omega$-$\lambda$ plane 
using the angular diameter distance of the FLRW universe. 
The dotted line connecting $(\Omega, \lambda)=(1,0)$ and $(0,1)$ 
indicates the spatially flat universe. 
(a)We assume $z_L=0.1$ and $z_S=4$. 
Each contour from top to bottom is drawn where $D_{LS}/D_{OS}$ runs 
from $0.97$ to $0.91$ at the interval of $0.01$. 
(b)We assume $z_L=0.3$ and $z_S=4$. 
Each contour from top to bottom is drawn where $D_{LS}/D_{OS}$ runs 
from $0.90$ to $0.76$ at the interval of $0.02$. 
(c)We assume $z_L=1$ and $z_S=4$. 
Each contour from top to bottom is drawn where $D_{LS}/D_{OS}$ runs 
from $0.64$ to $0.44$ at the interval of $0.02$. 
}

\figcaption[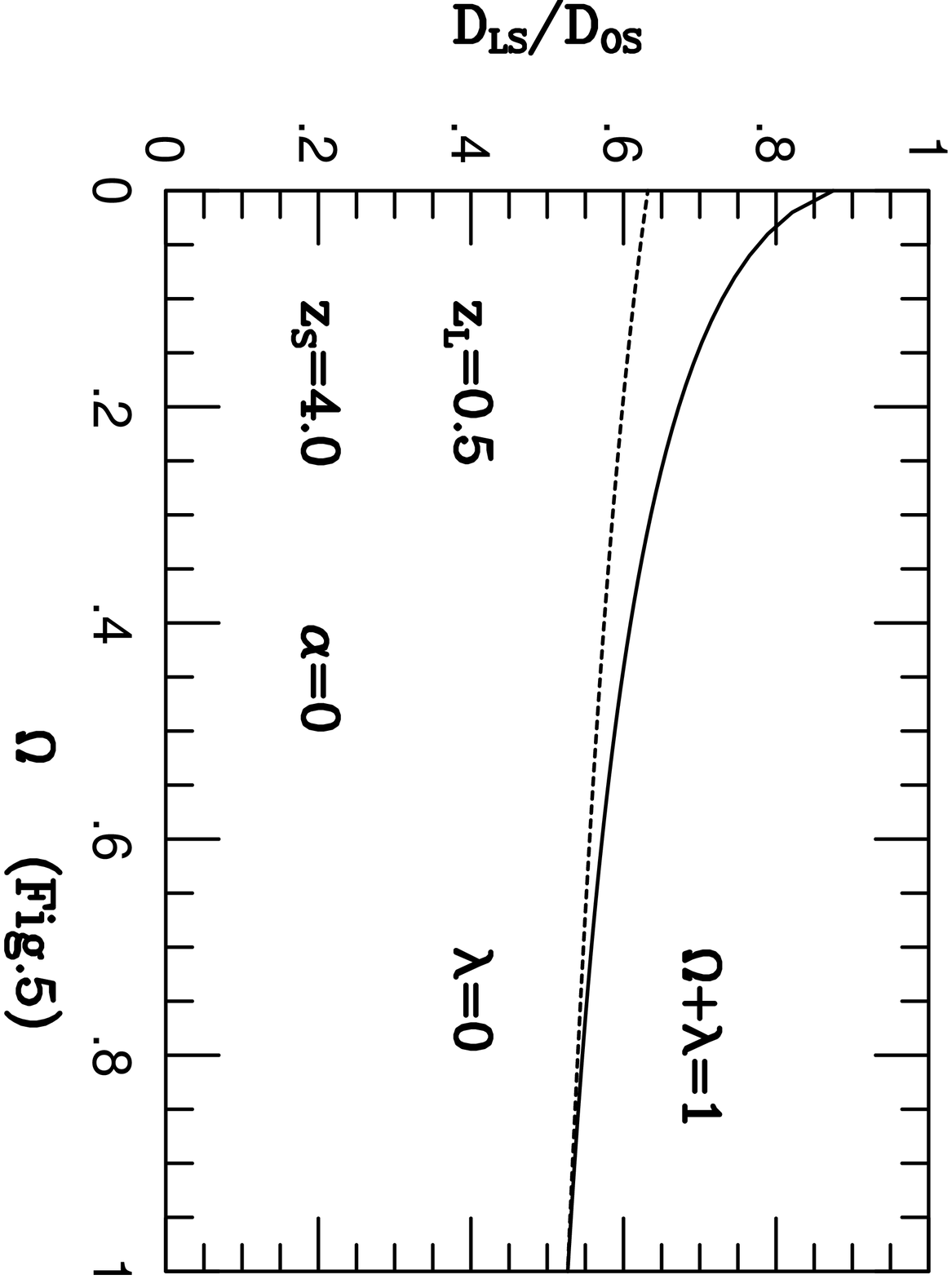]{
The distance ratio $D_{LS}/D_{OS}$ for the lens at $z_L=0.5$ 
and the source at $z_S=4$. 
We use DR angular diameter distance with $\alpha=0$. 
We assume $\Omega+\lambda=1$ (solid line) and $\lambda=0$ (dashed line). 
}

\figcaption[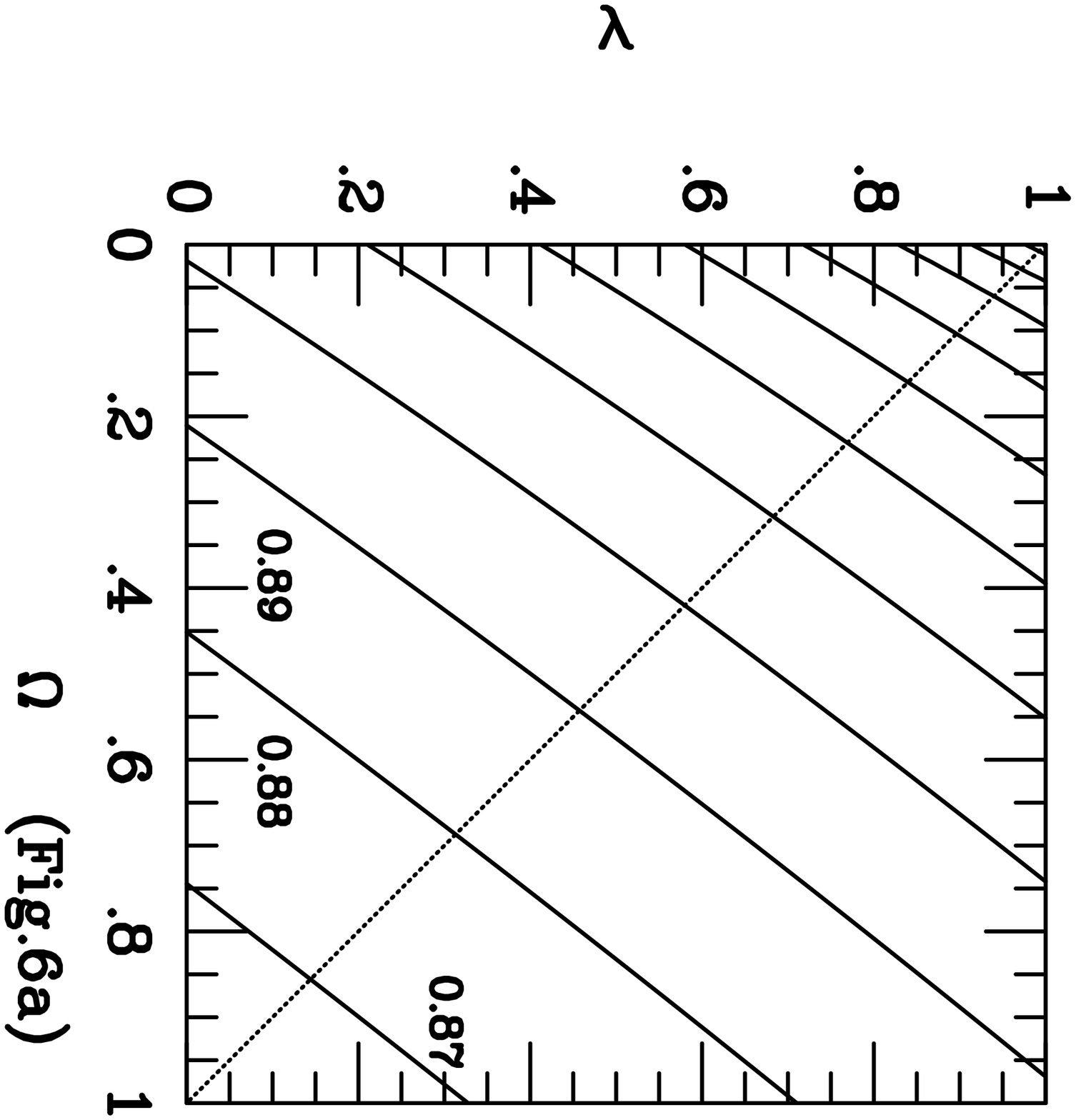,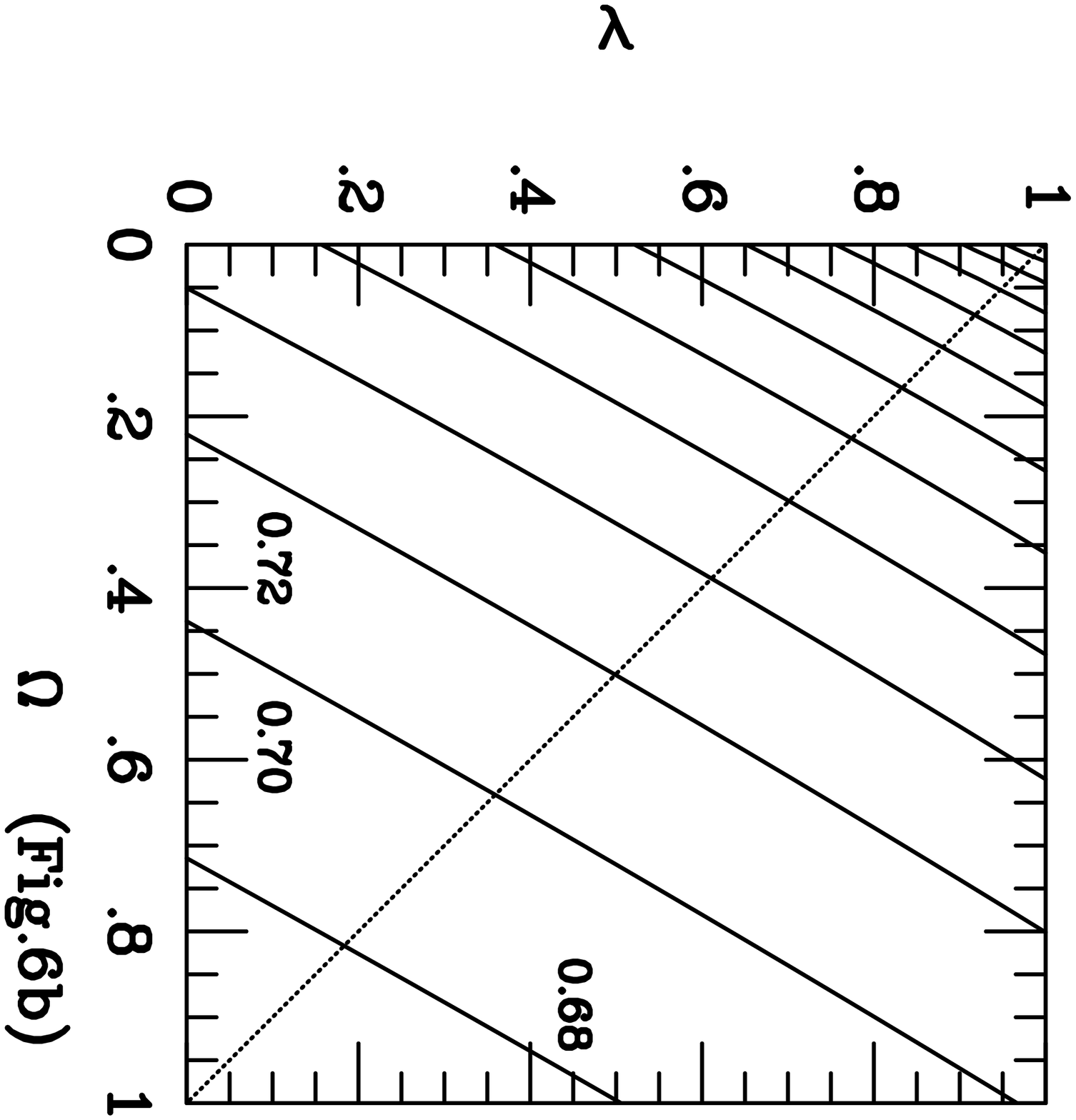,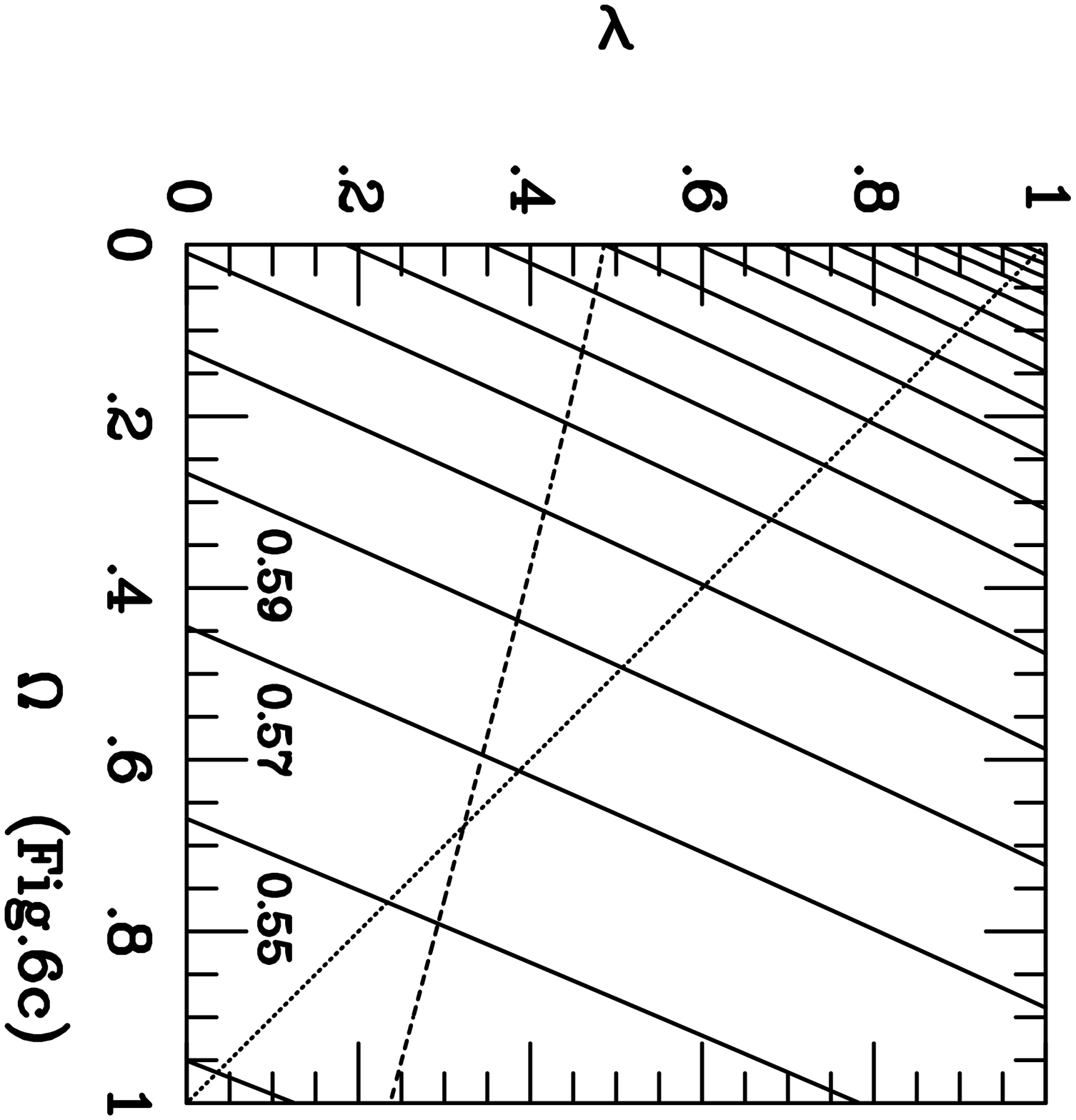,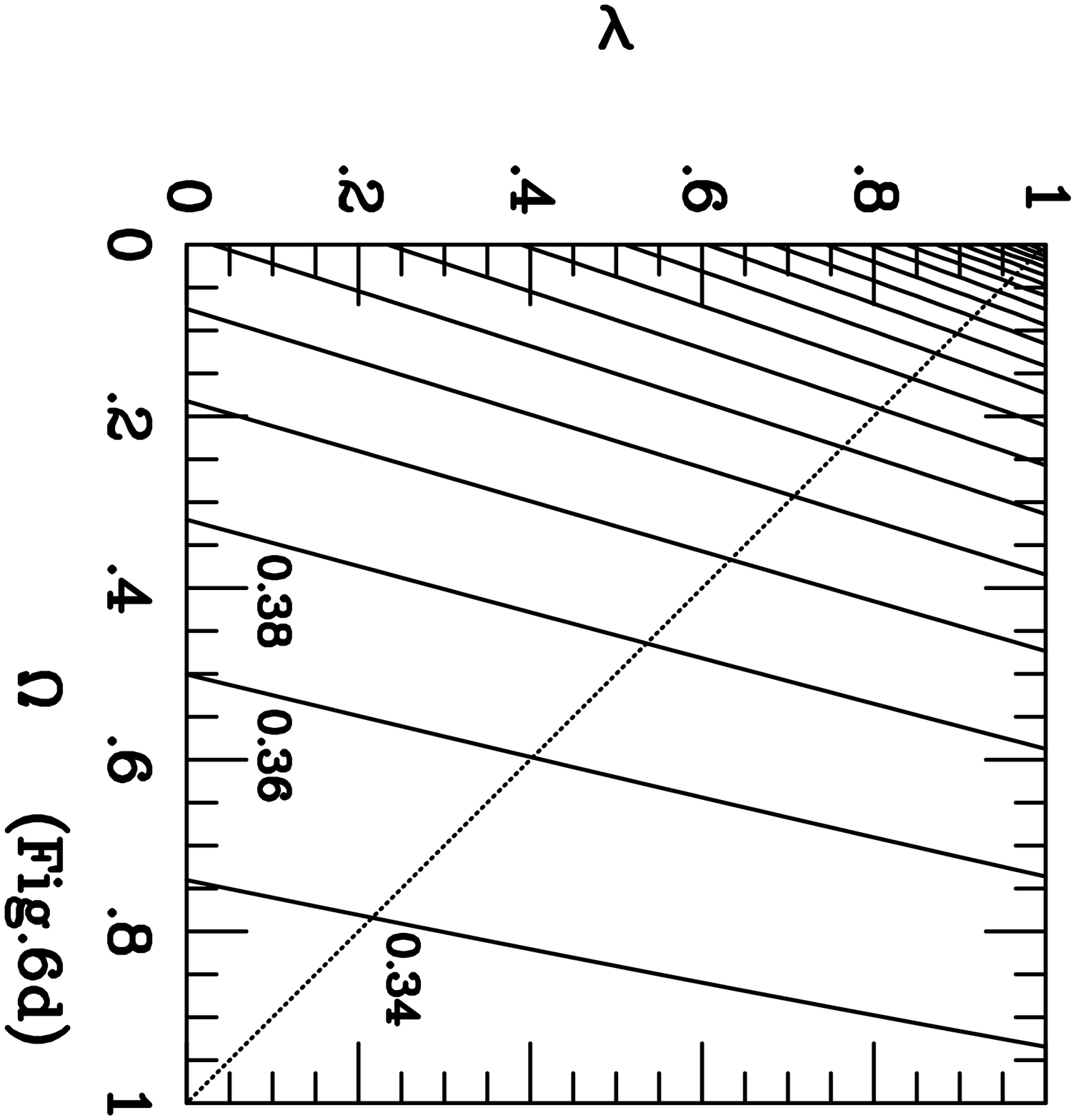]{
Contours of $D_{LS}/D_{OS}$ on $\Omega$-$\lambda$ plane, 
using DR angular diameter distance with $\alpha=0$. 
The dotted line connecting $(\Omega, \lambda)=(1,0)$ and $(0,1)$ 
indicates the spatially flat universe. 
(a)We assume $z_L=0.1$ and $z_S=4$. 
Each contour from top to bottom is drawn where $D_{LS}/D_{OS}$ runs 
from $0.97$ to $0.87$ at the interval of $0.01$. 
(b)We assume $z_L=0.3$ and $z_S=4$. 
Each contour from top to bottom is drawn where $D_{LS}/D_{OS}$ runs 
from $0.90$ to $0.68$ at the interval of $0.02$. 
(c)We assume $z_L=0.5$ and $z_S=4$. 
Each contour from top to bottom is drawn where $D_{LS}/D_{OS}$ runs 
from $0.85$ to $0.53$ at the interval of $0.02$. 
The dashed line is drawn at the level of $D_{LS}/D_{OS}=0.69$ for 
$\alpha=1$ for comparison. 
(d)We assume $z_L=1$ and $z_S=4$. 
Each contour from top to bottom is drawn where $D_{LS}/D_{OS}$ runs 
from $0.70$ to $0.34$ at the interval of $0.02$. 
}


\end{document}